\documentclass[twocolumn,journal]{IEEEtran}
\usepackage[latin9]{inputenc}
\usepackage{array}
\usepackage{float}
\usepackage{multirow}
\usepackage{amsmath}
\usepackage{amssymb}
\usepackage{graphicx}
\usepackage{color}
\usepackage[unicode=true,
 bookmarks=true,bookmarksnumbered=true,bookmarksopen=true,bookmarksopenlevel=1,
 breaklinks=false,pdfborder={0 0 0},pdfborderstyle={},backref=false,colorlinks=false]
 {hyperref}
\hypersetup{pdftitle={Your Title},
 pdfauthor={Your Name},
 pdfpagelayout=OneColumn, pdfnewwindow=true, pdfstartview=XYZ, plainpages=false}

\makeatletter

\providecommand{\tabularnewline}{\\}
\newcommand{\lyxdot}{.}

\floatstyle{ruled}
\newfloat{algorithm}{tbp}{loa}
\providecommand{\algorithmname}{Algorithm}
\floatname{algorithm}{\protect\algorithmname}

\let\oldforeign@language\foreign@language
\DeclareRobustCommand{\foreign@language}[1]{%
  \lowercase{\oldforeign@language{#1}}}

\usepackage[caption=false,font=footnotesize]{subfig}
\usepackage[noadjust]{cite}

\makeatother

\begin{document}
\title{Enhancing Diversity of OFDM with Joint Spread Spectrum and Subcarrier
Index Modulations}
\author{Vu-Duc Ngo,~Thien Van Luong, Nguyen Cong Luong, Mai Xuan Trang,
Minh-Tuan Le, Thi Thanh Huyen Le, and Xuan-Nam Tran\thanks{V.-D. Ngo is with the School of Electrical and Electronics Engineering,
Hanoi University of Science and Technology, Hanoi 11657, Vietnam,
(email: duc.ngovu@hust.edu.vn).}\thanks{T. V. Luong, N. C. Luong, M. X. Trang are with the Faculty of Computer
Science, Phenikaa University, Hanoi 12116, Vietnam (e-mail: \{thien.luongvan,
luong.nguyencong, trang.maixuan\}@phenikaa-uni.edu.vn).}\thanks{M.-T. Le is with the MobiFone R\&D Center, MobiFone Corporation, Hanoi
11312, Vietnam, (e-mail: tuan.minh@mobifone.vn).}\thanks{X.-N. Tran and T. T. H. Le are with the Advanced Wireless Communications
Group, Le Quy Don Technical University, Ha Noi 11355, Vietnam (e-mail:
\{namtx, huyen.ltt\}@mta.edu.vn).}}
\markboth{}{}
\maketitle
\begin{abstract}
This paper proposes a novel spread spectrum and sub-carrier index
modulation (SS-SIM) scheme, which is integrated to orthogonal frequency
division multiplexing (OFDM) framework to enhance the diversity over
the conventional IM schemes. Particularly, the resulting scheme, called
SS-SIM-OFDM, jointly employs both spread spectrum and sub-carrier
index modulations to form a precoding vector which is then used to
spread an $M$-ary complex symbol across all active sub-carriers.
As a result, the proposed scheme enables a novel transmission of three
signal domains: SS and sub-carrier indices, and a single $M$-ary
symbol. For practical implementations, two reduced-complexity near-optimal
detectors are proposed, which have complexities less depending on
the $M$-ary modulation size. Then, the bit error probability and
its upper bound are analyzed to gain an insight into the diversity
gain, which is shown to be strongly affected by the order of sub-carrier
indices. Based on this observation, we propose two novel sub-carrier
index mapping methods, which significantly increase the diversity
gain of SS-SIM-OFDM. Finally, simulation results show that our scheme
 achieves better error performance than the benchmarks at the cost
of lower spectral efficiency compared to classical OFDM and OFDM-IM,
which can carry multiple $M$-ary symbols.

\end{abstract}

\begin{IEEEkeywords}
OFDM-IM, SS-SIM, spread spectrum, index modulation, spreading, precoding,
Zadoff-Chu, detection designs.
\end{IEEEkeywords}

\section{Introduction}

\textcolor{black}{Index modulation (IM) \cite{SurveyIM} has recently emerged as a promising
modulation that exploits the indices of active sub-channels such as
antennas \cite{SM}, spreading codes \cite{Kaddoum2015} or sub-carriers
\cite{PCofdm1999} to carry additional data bits without requiring extra power or bandwidth. Therefore, IM not only exhibits higher reliability
and energy efficiency (EE), but also lower complexity than conventional
modulation schemes. Furthermore, it can enjoy an attractive trade-off between
the spectral efficiency (SE) and the EE just by adjusting the number
of active sub-channels. This makes IM a competitive candidate for
machine type communications (MTC) \cite{Bockelmann2016MTC}, which require a high flexibility
in terms of reliability, data rate and complexity.}

\textcolor{black}{The IM concept was first introduced to the orthogonal frequency division
multiplexing (OFDM) in \cite{PCofdm1999}, and was then comprehensively
analyzed in \cite{basar3013}. The resulting scheme, termed as OFDM-IM,
activates only a subset of sub-carriers to convey data bits via both
the active indices and conventional $M$-ary symbols. Subsequently,
a variety of IM techniques have been studied, which can be found in the
survey \cite{SurveyIM}. For example, the analyses of the bit error
rate (BER), achievable rate, and outage performance of OFDM-IM were 
presented in \cite{ThienTVT2017,rateIM,Pout2017}, respectively.  Additionally, a number of studies aimed to improve
the SE of OFDM-IM can be found in \cite{GeneralizedIM,dualmode,MultiModeIM,Wang2022Dual}, in which  extra bits are carried via additional signal domains, such as the index activation of inphase/quadrature (I/Q) components \cite{GeneralizedIM},  subcarriers' constellation modes \cite{dualmode,MultiModeIM} and 3-D constellation symbols \cite{Wang2022Dual}. In \cite{NoncoherentOFDMIM}, noncoherent OFDM-IM was proposed, which conveys information only via the indices of active antennas, and thus, its signal detection does not require any channel state information (CSI). Then, a number of techniques were proposed to  improve the SE of noncoherent OFDM-IM to be larger than 1 bps/Hz, such as deep learning (DL) \cite{Luong2020energy} and multi-level index modulation \cite{Fazeli2022Level}. Furthermore, the IM concept was  utilized in other multicarrier waveforms such as orthogonal time frequency space (OTFS) modulation \cite{Feng2022OTFS} and filter bank OFDM \cite{Huaijin2021Filter} in order to effectively combat Doppler effects in high-mobility channels and to reduce out-of-band emission (OOB), respectively. Recently, DL with deep neural networks (DNN) has also been applied to OFDM-IM as well as multicarrier systems in general. For instance,   
a deep learning-based detector for OFDM-IM called DeepIM was  proposed
in \cite{DeepIM2019}, which achieves a near-optimal performance at
even lower runtime complexity than the energy-based greedy detector
\cite{ThienTVT2017}, followed by its reduced complexity version based on convolutional neural network (CNN) \cite{Wang2020Conv}. More specifically, a DNN architecture called autoencoder was developed for jointing representing and optimizing both transmitter and receiver of optical/wireless multicarrier OFDM-based systems in \cite{Luong2022optical,Zhang2022Machine,Luong2021mcae,Chao2022tubro}. In other context, in \cite{mimoIMbasar2016}, OFDM-IM is combined with multi-input multi-output
(MIMO) systems, while in \cite{JamesTVT,thienWCL2018} it is combined
with both diversity receptions and greedy detection. Next, we will discuss the methods for improving the diversity of OFDM-IM without relying additional hardware such as antennas.}

\textcolor{black}{As inherited from classical OFDM, the performance of OFDM-IM is poor
under a fading channel environment. To improve the reliability, a
range of transmit diversity schemes have been proposed for OFDM-IM,
without using multiple antennas. For instance, in \cite{CIbasar2015},
coordinate interleaved OFDM-IM (CI-OFDM-IM) was proposed to transmit
real and imaginary parts of $M$-ary symbols via different sub-carriers,
achieving a diversity gain of two. In \cite{EnhancedIM2017}, a linear
constellation precoder was designed for OFDM-IM, also providing a diversity
order of two. The repeated multicarrier index keying OFDM (MCIK-OFDM)
was proposed in \cite{ThienTWC2018} to carry the same $M$-ary symbol
over all active sub-carriers. The resulting scheme, termed as ReMO,
increases the diversity gain up to two at the expense of the SE. Note
that the terms of MCIK-OFDM and OFDM-IM are equivalent. Meanwhile,
the repetition code for sub-carrier index (SI) symbols was presented
in \cite{codedIM2017choi}, combining with channel coding for $M$-ary
symbols. A similar index repetition method for CI-OFDM-IM was reported
in \cite{Le2020repeated}. In \cite{SSIM2018}, IM was applied to spread
spectrum OFDM (SS-OFDM), where the indices of spreading codes are
utilized to carry extra data bits and the resulting scheme is called SS-OFDM-IM.
The spread matrices were applied to OFDM-IM in \cite{ThienTVT2018}
to enhance its diversity gain at the cost of higher receiver complexity. In \cite{Chao2021Unified}, the discrete Fourier transform (DFT) was applied to OFDM-IM to either improve its error performance or reduce its peak-to-average power ratio (PAPR) as well as OOB.}

Note that most of the aforementioned schemes exploit only two signal
dimensions including $M$-ary symbols and either sub-carrier indices
or spreading code indices to offer the diversity gain limited by two.
In this work, we propose a novel IM scheme which exploits all possible
three IM signal dimensions, while achieving the diversity gain of
even more than two. Our contributions are summarized as follows.
\begin{itemize}
\item A novel spread spectrum and sub-carrier index modulation for OFDM
(SS-SIM-OFDM) is proposed, which jointly exploits SS and SI modulations
to create a precoding matrix, which is then used to spread an $M$-ary
complex symbol across all active sub-carriers. Interestingly, SS-SIM-OFDM
not only forms a new transmission of three domains including SS and
sub-carrier indices, and $M$-ary symbols, but also improves diversity
gain over existing IM schemes with two signal domains \cite{ThienTWC2018,SSIM2018}.
\item We design two low-complexity detectors, namely the near-maximum likelihood
(near-ML) and log-likelihood ratio-maximal ratio combining (LLR-MRC)
detectors, whose complexities are shown to be less depending on the
$M$-ary modulation size. Particularly, LLR-MRC separately detects
sub-carrier indices and other signal domains, thus has much lower
complexity than the near-ML detector at the cost of a slight performance
loss.
\item Based on the bit error probability (BEP) analysis, we discover that
unlike existing IM schemes, the diversity gain of the proposed scheme
is strongly influenced by the order of sub-carrier indices in each
active SI set. This prompts us to propose two novel SI mapping methods
to significantly enhance diversity gain of SS-SIM-OFDM.
\item Simulations are carried out to verify the superior BEP of SS-SIM-OFDM
over benchmark schemes, as well as the effectiveness of proposed diversity
enhancement methods and reduced-complexity detectors.
\end{itemize}
The rest of this paper is organized as follows. Section II presents
the system model, while Section III proposes two low-complexity detectors.
The BEP analysis and diversity enhancement are reported in Section
IV. In Section V, simulation results are provided. Finally, Section
VI concludes the paper.

\textit{Notation:} Upper-case bold and lower-case bold letters present
matrices and vectors, respectively. $(.)^{T}$ and $(.)^{H}$ stand
for the transpose and Hermitian operations, respectively. $\left\Vert .\right\Vert $
denotes the Frobenious norm. $\mathcal{CN}\left(0,\sigma^{2}\right)$
represents the complex Gaussian distribution with zero mean and variance
$\sigma^{2}$. The ring of complex number, the binomial coefficient
and the floor function are denoted by $\mathbb{C},$ $C\left(.,.\right)$
and $\left\lfloor .\right\rfloor $, respectively. $j$ is the unit
imaginary number. The greatest common divisor of two integers is denoted
by $\text{gcd}\left(.,.\right)$.

\section{System Model}

\begin{figure}[tb]
\begin{centering}
\includegraphics[width=0.9\columnwidth]{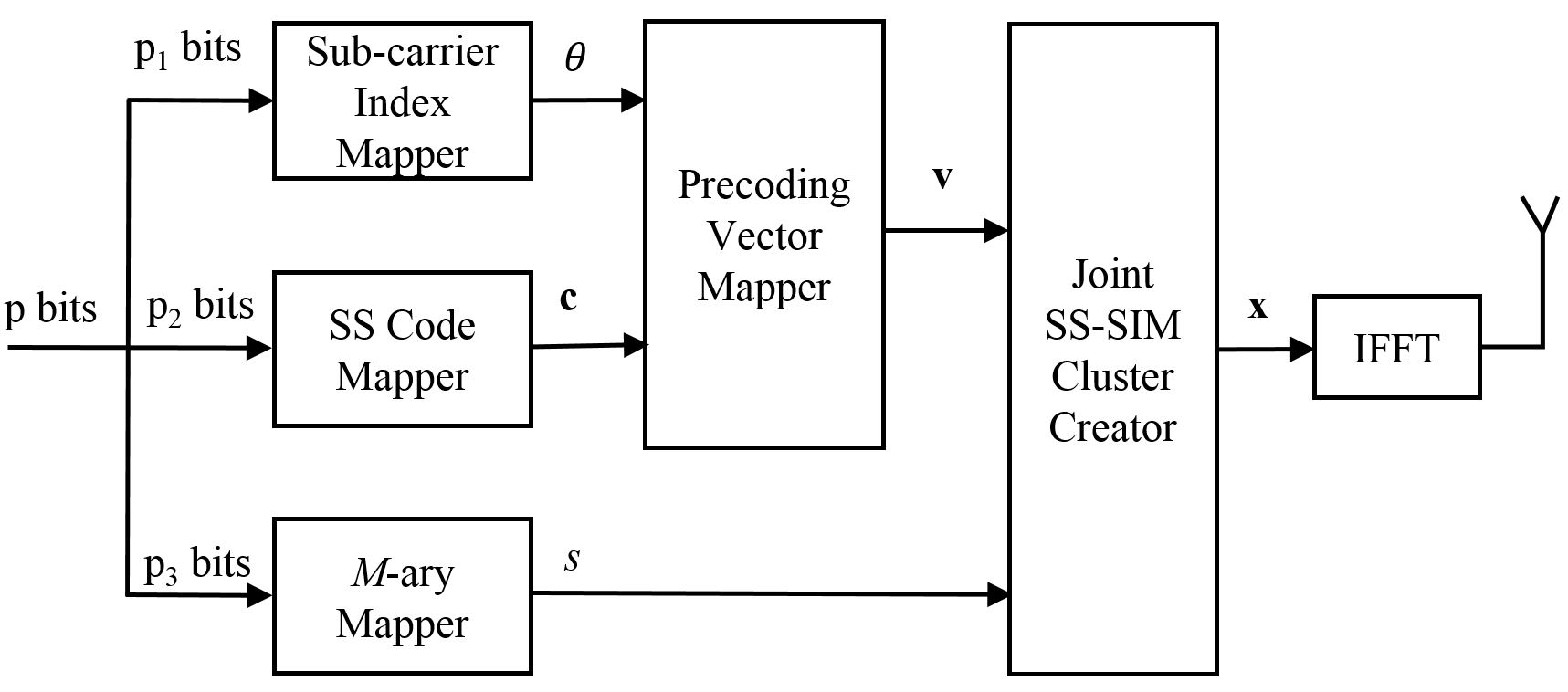}
\par\end{centering}
\caption{Block diagram of one cluster of SS-SIM-OFDM. \label{fig:SpeadIM_diagram}}
\end{figure}

Consider an SS-SIM-OFDM with $N_{c}$ sub-carriers, which are divided
into $G$ clusters and each cluster has $N$ sub-carriers, where $N_{c}=GN$.
Without loss of generality, we only consider one cluster for simplicity.
A block diagram of one SS-SIM-OFDM cluster is illustrated in Fig.
\ref{fig:SpeadIM_diagram}. Unlike exiting IM schemes which carry
information mainly via only two signal dimensions \cite{SurveyIM},
the proposed SS-SIM-OFDM activates $K$ out of $N$ sub-carriers in
every transmission per cluster to convey data bits through three signal
domains as follows.

For each transmission in one cluster, $p$ incoming bits are partitioned
into three bit streams of $p_{1}$, $p_{2}$ and $p_{3}$ bits, where
$p=p_{1}+p_{2}+p_{3}$. The first stream of $p_{1}$ bits are mapped
into the set of $K$ indices of active sub-carriers, which is denoted
by $\theta=\left\{ \alpha_{1},...,\alpha_{K}\right\} $, where $\alpha_{k}\in\left\{ 1,2,...,N\right\} $
for $k=1,...,K$. Here, $\theta$ is referred to as a sub-carrier
index (SI) symbol. The second stream of $p_{2}$ bits are fed to the
SS code mapper to determine the spreading code $\mathbf{c}=\left[c_{1},...,c_{K}\right]^{T}\in\mathcal{C}\subseteq\mathbb{C}^{K\times1}$,
where $\mathcal{C}=\left\{ \mathbf{c}_{1},...,\mathbf{c}_{K}\right\} $
is the set of $K$ orthogonal codes.\footnote{Note that since we aim to spread the $M$-ary symbol over $K$ active
sub-carriers, the length of the spreading codes is equal to $K$.} The design of spreading codes will be discussed afterwards. Then,
the precoding vector mapper utilizes $\theta$ and $\mathbf{c}$ to
form the precoding vector $\mathbf{v}=\left[v_{1},...,v_{N}\right]^{T}$,
where $v_{\alpha}=c_{k}$ for $\alpha=\alpha_{k}\in\theta$, $k=1,...,K$
and $v_{\alpha}=0$ for $\alpha\notin\theta$. The precoding vector
mapper is denoted by the function $\mathbf{v}=\mathcal{T}\left(\theta,\mathbf{c}\right)$.
For example, when $N=4$, $K=2$ and $\theta=\left\{ 2,3\right\} $,
we obtain $\mathbf{v}=\left[0,c_{1},c_{2},0\right]^{T}$. The remaining
$p_{3}$ bits are mapped to an $M$-ary data symbol $s\in\mathcal{S}$,
where $\mathcal{S}$ denotes the $M$-ary modulation constellation.
Finally, the transmitted signal in the frequency domain for each cluster
is obtained by spreading this symbol across $K$ active sub-carriers
as follows $\mathbf{x}=\mathbf{v}s$.

The numbers of bits carried by sub-carrier and spreading code indices
are $p_{1}=\left\lfloor \log_{2}C\left(N,K\right)\right\rfloor $
and $p_{2}=\left\lfloor \log_{2}K\right\rfloor $, respectively, while
that of $M$-ary bits is $p_{3}=\log_{2}M$. Thus, the SE of SS-SIM-OFDM
is given by
\begin{equation}
R=\frac{\left\lfloor \log_{2}C\left(N,K\right)\right\rfloor +\left\lfloor \log_{2}K\right\rfloor +\log_{2}M}{N}\,\,\text{(bps/Hz).}\label{eq:Rate}
\end{equation}

The received signal in the frequency domain is given by
\begin{equation}
\mathbf{y}=\mathbf{Hv}s+\mathbf{n}=\mathbf{Hx}+\mathbf{n},\label{eq:y}
\end{equation}
where $\mathbf{H}=\text{diag}\left(h_{1},...,h_{N}\right)$ denotes
the Rayleigh fading channel matrix with $h_{n}\sim\mathcal{CN}\left(0,1\right)$
and $\mathbf{n}\in\mathbb{C}^{N\times1}$ is the additive white Gaussian
noise vector with its elements $\sim\mathcal{CN}\left(0,N_{0}\right)$.
Here, we assume the average receive signal-to-noise ratio (SNR) is
$\bar{\gamma}=1/N_{0}$. At the receiver, the transmitted signal can
be detected by the maximum likelihood (ML) detector as follows:
\begin{equation}
\left(\hat{\theta},\hat{\mathbf{c}},\hat{s}\right)=\arg\min_{\theta,\mathbf{c},s}\left\Vert \mathbf{y}-\mathbf{Hv}s\right\Vert ^{2}.\label{eq:ML_detector}
\end{equation}
This detector has a computational complexity of $\mathcal{O}\left(2^{p}\right)$,
which is impractical when $N$, $K$ and $M$ increase. Therefore,
in Subsection III, we will design two low-complexity, near-optimal
detectors for our SS-SIM-OFDM.

It is noteworthy that SS-SIM-OFDM jointly employs both SS and SI modulations,
enabling a novel transmission of three signal domains, including the
indices of subcarriers and spreading code, and the $M$-ary symbol.
Hence, our scheme provides a higher SE than the IM schemes based on
two signal domains only, such as ReMO \cite{ThienTWC2018} and SS-OFDM-IM
\cite{SSIM2018}, which transmit only a single $M$-ary symbol and
an index symbol in either SI or SS domain. However, compared to classical
OFDM and OFDM-IM, our scheme achieve lower SE for given values of
$N$, $K$ and $M$, since these classical schemes can carry multiple
$M$-ary symbols.\footnote{In our future work, we will attempt to increase the SE of the proposed
scheme through conveying multiple $M$-ary symbols similar to OFDM-IM,
while still ensuring the benefit of three signal domains.} \textcolor{black}{Additionally, the diversity gain of SS-SIM-OFDM can be notably enhanced,
even to be higher than that of the existing IM schemes. This can be achieved through properly designing
the SI mapper so that each $M$-ary data symbol or index symbol is conveyed involving as much different sub-carriers as possible, as will be shown in Section IV.}

\subsection{Designs of Spreading Codes }

There are a number of well-known orthogonal spreading codes that can
be used for SS-SIM-OFDM such as Walsh-Hadarmard (WH) and Zadoff-Chu
(ZC) codes. Notice from \cite{SSIM2018} that ZC not only offers
better performance, but also has more flexible length than WH. More
precisely, the length of ZC can be any positive integer, while that
of WH must be an integer power of two. Consequently, we consider ZC
for the proposed scheme. 

In particular, the first ZC code is denoted by $\mathbf{c}_{1}=\left[c_{1},...,c_{K}\right]^{T}$
which is given by
\begin{equation}
c_{k}=\begin{cases}
e^{-\frac{j2\pi d}{K}\left(\frac{k^{2}}{2}+uk\right)} & \text{for}\text{\,\,even\,\ensuremath{\,K}}\\
e^{-\frac{j2\pi d}{K}\left[\frac{k\left(k+1\right)}{2}+uk\right]} & \text{for}\,\,\text{odd\,}\,K
\end{cases},\label{eq:ZC_sequence}
\end{equation}
where $d$ is any integer relatively prime to $K$, $u$ is any integer
and $k=1,...,K$. Then, the $k$-th ZC code is determined as the $k$-th
cyclically shifted version of $\mathbf{c}_{1}$. For instance, $\mathbf{c}_{2}=\left[c_{K},c_{1},...,c_{K-1}\right]^{T}$
and $\mathbf{c}_{3}=\left[c_{K-1},c_{K},...,c_{K-2}\right]^{T}$.
To further improve the performance, we propose to use the rotated
ZC spreading codes which are defined by rotating each original ZC
code $\mathbf{c}_{k}$ by a dedicated angle of $e^{\frac{j2\pi\left(k-1\right)}{B}}$
for $k=1,...,K$, where $B=MK-1$. It is worth noting that we choose
$B=MK-1$ to maximize the diversity gain achieved in the SS index
domain. This will be explained further in the performance analysis
afterwards in Subsection IV.

\section{Low Complexity Receiver Designs}

We propose two low-complexity detectors for SS-SIM-OFDM, namely the
near-ML and log-likelihood ratio maximal ratio combining (LLR-MRC)
detectors. The computational complexity analysis and comparison are
also provided.

\subsection{Near-ML Detector}

Let us denote the set of SI symbols used for the SIM process as $\mathcal{I}=\left\{ \theta_{1},...,\theta_{2^{p_{1}}}\right\} .$
The near-ML detector first fixes the SI symbol $\theta_{i}\in\mathcal{I}$
to detect the corresponding spreading code $\mathbf{\hat{c}}_{i}$
and $M$-ary symbol $\hat{s}_{i}$ as follows. For a given vector
$\mathbf{u}=\left[u_{1},...u_{N}\right]^{T}\in\mathbb{C}^{N\times1},$
denote by $\mathbf{u}\left(\theta\right)$ the $K\times1$ vector
whose elements have the indices of $\theta$. For example, when $(N,K)=(4,2)$
and $\theta=\{1,3\}$, we attain $\mathbf{u}\left(\theta\right)=\left[u_{1},u_{3}\right]^{T}$.
The same operation can be applied to the diagonal matrix such as $\mathbf{U}\left(\theta\right)=\text{diag\ensuremath{\left(u_{1},u_{3}\right)},}$
where $\mathbf{U}=\text{diag\ensuremath{\left(\mathbf{u}\right)}}$.
Based on this definition, for each $\theta_{i}\in\mathcal{I}$, we
extract $\mathbf{H}_{i}=\mathbf{H}\left(\theta_{i}\right)$ and $\mathbf{y}_{i}=\mathbf{y}\left(\theta_{i}\right).$
Then, for each $\mathbf{c}_{k}\in\mathcal{C}$ with $k=1,...,2^{p_{2}}$,
the $M$-ary symbol $s_{i,k}$ is estimated based on the MRC as 
\begin{equation}
s_{i,k}=\mathcal{Q}\left\{ \mathbf{H}_{i,k}^{H}\mathbf{y}_{i}/W_{i}\right\} ,\label{eq:s_ik}
\end{equation}
where $\mathbf{H}_{i,k}=\mathbf{H}_{i}\mathbf{c}_{k}$, $W_{i}=\left\Vert \mathbf{H}_{i}\right\Vert ^{2}$
and $\mathcal{Q}\left(s\right)$ represents the digital demodulator
function that returns the $M$-ary symbol which is the closest one
to $s$. The symbol $s_{i,k}$ is then utilized to calculate the distance
$\Delta_{i,k}=\left\Vert \mathbf{y}_{i}-\mathbf{H}_{i,k}s_{i,k}\right\Vert ^{2}.$
After obtaining $2^{p_{2}}$ symbols $s_{i,k}$ and distances $\Delta_{i,k}$,
we can generate the best $\mathbf{\hat{c}}_{i}$ and $\hat{s}_{i}$
for each $\theta_{i}$, as follows
\begin{equation}
\mathbf{\hat{c}}_{i}=\mathbf{c}_{\hat{k}_{i}},\,\,\hat{s}_{i}=s_{i,\hat{k}_{i}},\label{eq:ci-si}
\end{equation}
where $\hat{k}_{i}=\arg\min_{k=1,...,2^{p_{2}}}\Delta_{i,k}$. We
use \eqref{eq:ci-si} to compute the overall distance $\Theta_{i}=\left\Vert \mathbf{y}-\mathbf{H}\mathbf{v}_{i}\hat{s}_{i}\right\Vert ^{2}$,
where $\mathbf{v}_{i}=\mathcal{T}\left(\theta_{i},\mathbf{\hat{c}}_{i}\right)$. 

Finally, the transmitted signal is recovered based on the minimum
distance out of $2^{p_{1}}$ distances $\Theta_{i}$, as follows
\begin{equation}
\hat{\theta}=\theta_{\hat{i}},\,\,\hat{\mathbf{c}}=\mathbf{\hat{c}}_{\hat{i}},\,\,\hat{s}=\hat{s}_{\hat{i}},\label{eq:final_1}
\end{equation}
where $\hat{i}=\arg\min_{i=1,...,2^{p_{1}}}\Theta_{i}$. 

\textcolor{black}{The proposed
near-ML detector is summarized in Algorithm 1. Its main idea is that for given sub-carrier activation $\theta_i$, it extracts the corresponding active sub-channels $\mathbf{H}_{i}$ in order to detect the corresponding spreading code and data symbol based on the minimum Euclidean distance criterion.}

Since the near-ML detector only uses sub-channels, i.e., $\mathbf{H}_{i}$
and the MRC for detection of the spreading code and the $M$-ary symbol,
its complexity can be considerably reduced compared to the ML in \eqref{eq:ML_detector}.
Despite that, our near-ML still achieves the near-ML performance,
as shown in Section V.

\begin{algorithm}[tbh]
\caption{Near-ML Detection Algorithm}
\label{alg:Near-ML} \textbf{Input:} $\mathbf{y}$, $\mathbf{H}$,
$\mathcal{C}$ and $\mathcal{I}$

\textbf{Output:} $\hat{\theta}$, $\hat{\mathbf{c}}$ and $\hat{s}$ 
\begin{enumerate}
\item \textbf{for} $i=1$ \textbf{to} $2^{p_{1}}$ \textbf{do}
\item ~~~Extract $\mathbf{H}_{i}=\mathbf{H}\left(\theta_{i}\right)$
and $\mathbf{y}_{i}=\mathbf{y}\left(\theta_{i}\right),$ where $\theta_{i}\in\mathcal{I}$.
\item ~~~Calculate $W_{i}=\left\Vert \mathbf{H}_{i}\right\Vert ^{2}$.
\item ~~~\textbf{for} $k=1$ \textbf{to} $2^{p_{2}}$ \textbf{do}
\item ~~~~~~Calculate $\mathbf{H}_{i,k}=\mathbf{H}_{i}\mathbf{c}_{k}$,
where $\mathbf{c}_{k}\in\mathcal{C}$.
\item ~~~~~~Estimate $s_{i,k}=\mathcal{Q}\left\{ \mathbf{H}_{i,k}^{H}\mathbf{y}_{i}/W_{i}\right\} \in\mathcal{S}$.
\item ~~~~~~Compute $\Delta_{i,k}=\left\Vert \mathbf{y}_{i}-\mathbf{H}_{i,k}s_{i,k}\right\Vert ^{2}.$
\item ~~~\textbf{end for}
\item ~~~Estimate $\hat{k}_{i}=\arg\min_{k=1,...,2^{p_{2}}}\Delta_{i,k}$.
\item ~~~Generate $\mathbf{\hat{c}}_{i}=\mathbf{c}_{\hat{k}_{i}}$ and
$\hat{s}_{i}=s_{i,\hat{k}_{i}}$.
\item ~~~Compute $\Theta_{i}=\left\Vert \mathbf{y}-\mathbf{H}\mathbf{v}_{i}\hat{s}_{i}\right\Vert ^{2}$,
where $\mathbf{v}_{i}=\mathcal{T}\left(\theta_{i},\mathbf{\hat{c}}_{i}\right)$.
\item \textbf{end for}
\item Estimate $\ensuremath{\hat{i}=\arg\min_{i=1,...,2^{p_{1}}}\Theta_{i}}$.
\item Generate the output $\hat{\theta}=\theta_{\hat{i}}$, $\hat{\mathbf{c}}=\mathbf{\hat{c}}_{\hat{i}}$
and $\hat{s}=\hat{s}_{\hat{i}}$. 
\end{enumerate}
\end{algorithm}

\subsection{LLR-MRC Detector}

In order to further reduce the complexity of the near-ML detector,
we now propose the LLR-MRC detector which separately detects sub-carrier
indices, and the symbols in the two remaining signal dimensions. Particularly,
the LLR method is used to recover the indices of active sub-carriers
first, and then both the spreading code and the $M$-ary symbol are
estimated based on the MRC approach. 

In the first step, for each sub-carrier $n$, the LLR \cite{codedIM2017choi}
is computed by
\begin{equation}
\ensuremath{\lambda_{n}=-\min_{m=1,...,M}r_{m}}+\left|y_{n}\right|^{2},\label{eq:lamda_n}
\end{equation}
where $r_{m}=\left|y_{n}-h_{n}x_{m}\right|^{2}$ with $x_{m}\in\mathcal{S}$,
and $y_{n}$ is the $n$-th element of $\mathbf{y}$. Notice that
this ratio gives information about the activity status of the sub-carrier,
i.e., the larger LLR means the corresponding sub-carrier is more likely
to be active. Hence, the $N$ LLRs are arranged in the descending
order $\lambda_{\alpha_{1}}\ge...\ge\lambda_{\alpha_{N}},$ where
$\alpha_{n}\in\{1,...,N\}$ to decide $\hat{\theta}=\left\{ \alpha_{1},...,\alpha_{K}\right\} ,$
i.e., indices of $K$ largest LLR values. However, such an SI detection
may lead to the unexpected case of $\hat{\theta}\notin\mathcal{I}$.
To tackle this, we can replace $\alpha_{K}$ (the index of the smallest
LLR in $\hat{\theta}$) with $\alpha_{K+1}$ (the index of the largest
LLR not in $\hat{\theta}$) to obtain new indices as $\hat{\theta}=\left\{ \alpha_{1},...,\alpha_{K-1},\alpha_{K+1}\right\} $.
In case the new $\hat{\theta}$ is still not included in $\mathcal{I}$,
we continue to replace $\alpha_{K-1}\in\hat{\theta}$ with $\alpha_{K+2}\notin\hat{\theta}$.
In fact, such the case rarely occurs, carrying out the replacement
once is sufficient to improve the performance.

Next, ultizing $\hat{\theta}$, the MRC with the input of $\mathbf{\hat{H}}=\mathbf{H}\left(\hat{\theta}\right)$
and $\mathbf{\hat{y}}=\mathbf{y}\left(\hat{\theta}\right)$ is employed
to decode both $\hat{\mathbf{c}}$ and $\hat{s}$. This step is similar
to the inner for-loop of Algorithm 1. Thus, it is omitted for brevity.
The summary of LLR-MRC is given in Algorithm 2. It is worth noting
that the LLR-MRC exhibits remarkably lower complexity than the near-ML detector at the cost of the performance loss, as analyzed in the next
section. 

\begin{algorithm}[tbh]
\caption{LLR-MRC Detection Algorithm}
\label{alg:LLR-MRC} \textbf{Input:} $\mathbf{y}$, $\mathbf{H}$,
$\mathcal{C}$ and $\mathcal{I}$

\textbf{Output:} $\hat{\theta}$, $\hat{\mathbf{c}}$ and $\hat{s}$ 
\begin{enumerate}
\item \textbf{for} $n=1$ \textbf{to} $N$ \textbf{do}
\item ~~~Compute $r_{m}=\left|y_{n}-h_{n}x_{m}\right|^{2}$, $x_{m}\in\mathcal{S}$,
$m=1,..,M.$
\item ~~~Calculate LLR ratio $\ensuremath{\lambda_{n}=-\min_{m=1,...,M}r_{m}}+\left|y_{n}\right|^{2}$.
\item \textbf{end for}
\item Arrange $N$ ratios in descending order $\lambda_{\alpha_{1}}\ge...\ge\lambda_{\alpha_{N}},$
where $\alpha_{n}\in\{1,...,N\}$ to obtain $\hat{\theta}=\left\{ \alpha_{1},...,\alpha_{K}\right\} $.
\item \textbf{if} $\hat{\theta}\notin\mathcal{I}$
\item ~~~Update new indices $\hat{\theta}=\left\{ \alpha_{1},...,\alpha_{K-1},\alpha_{K+1}\right\} $.
\item \textbf{end if}
\item Extract $\mathbf{\hat{H}}=\mathbf{H}\left(\hat{\theta}\right)$ and
$\mathbf{\hat{y}}=\mathbf{y}\left(\hat{\theta}\right)$.
\item Calculate $W=\left\Vert \mathbf{\hat{H}}\right\Vert ^{2}.$
\item \textbf{for} $k=1$ \textbf{to} $2^{p_{2}}$ \textbf{do}
\item ~~~Calculate $\mathbf{\hat{H}}_{k}=\mathbf{\hat{H}}\mathbf{c}_{k}$,
where $\mathbf{c}_{k}\in\mathcal{C}$.
\item ~~~Estimate $s_{k}=\mathcal{Q}\left\{ \mathbf{\hat{H}}_{k}^{H}\mathbf{\hat{y}}/W\right\} \in\mathcal{S}$.
\item ~~~Compute $\Delta_{k}=\left\Vert \mathbf{\hat{y}}-\mathbf{\hat{H}}_{k}s_{k}\right\Vert ^{2}.$
\item \textbf{end for}
\item Estimate $\ensuremath{\hat{k}=\arg\min_{k=1,...,2^{p_{2}}}\Delta_{k}}$.
\item Generate the output $\hat{s}=s_{\hat{k}}$ and $\hat{\mathbf{c}}=\mathbf{c}_{\hat{k}}$. 
\end{enumerate}
\end{algorithm}

\subsection{Complexity Analysis and Comparison}

We evaluate the computational complexity of the near-ML, LLR-MRC and
ML detectors by calculating the number of floating-point operations
(flops) per sub-carrier. A flop can be a real addition, subtraction,
division or multiplication. For example, a complex multiplication
can be counted by 6 flops since it requires 2 real additions and 4
real multiplications, or the operation of $\left|z\right|^{2}$ where
$z\in\mathbb{C}$ requires 3 flops corresponding to 2 real multiplications
and 1 real addition.

Based on the definition of a flop, the numbers of flops required in
steps 3, 5, 6, 7 and 11 of the near-ML detector are $4K-1\approx4K$,
$6K$, $8K$, $12K$ and $14K+5N$, respectively. Notice that the
complexity involved in the digital demodulator function $\mathcal{Q}\left(.\right)$
is negligible in comparison with other calculations. As a result,
the number of flops per sub-carrier of the near-ML detector can be
approximated by
\begin{equation}
C_{\text{near-ML}}\approx\frac{2^{p_{1}}\left[\left(26\times2^{p_{2}}+18\right)K+4N\right]}{N}.\label{eq:C_nearML}
\end{equation}
Similarly, we attain the complexities of the LLR-MRC and ML detectors
in terms of flops/sub-carrier, respectively, as follows
\begin{equation}
C_{\text{LLR-MRC}}\approx\frac{2^{p_{2}}\times26K+4K}{N}+15,\label{eq:C_LLR_MRC}
\end{equation}

\begin{equation}
C_{\text{ML}}\approx\frac{2^{p_{1}+p_{2}}\left(4N+14K\right)M}{N}.\label{eq:C_ML}
\end{equation}

\begin{figure}[tb]
\begin{centering}
\includegraphics[width=1\columnwidth]{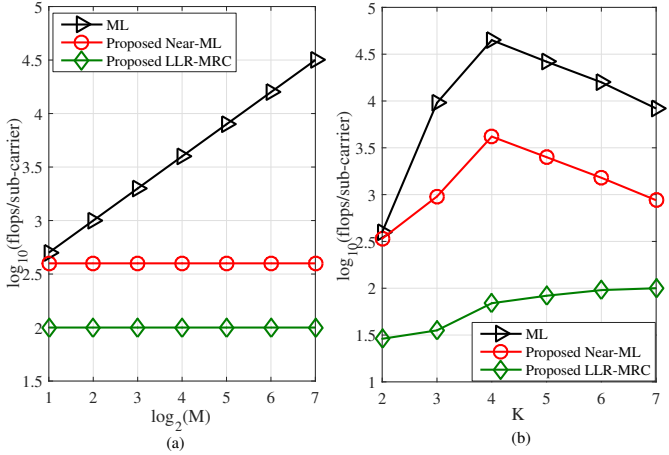} 
\par\end{centering}
\caption{Computational complexity comparisons between two proposed detectors
and the ML detector when (a) $N=5$, $K=4$, $M=2,4,...,128$ and
(b) $N=8,$ $M=16,$ $K=2,...,7$. \label{fig:Com}}
\end{figure}

It is worth mentioning that unlike the ML, the complexities of both
the proposed detectors are roughly independent of the $M$-ary modulation
size, thus are significantly lower than that of the ML, especially
when $M$ increases. For this, we provide Fig. \ref{fig:Com}(a) and
(b) to illustrate such a complexity reduction. It is clearly shown
from Fig. \ref{fig:Com} that the proposed detectors offer the notably
reduced complexity compared with the ML, especially the LLR-MRC detector.
For instance, in Fig. \ref{fig:Com}(a), when $(N,K,M)=(5,4,64)$,
the number of flops per sub-carrier of the ML detector is 15564, while
that of the proposed near-ML and LLR-MRC detectors are 406 and 101,
respectively. As a result, the near-ML and LLR-MRC detectors can save
$97.4\%$ and $99.3\%$ complexity, respectively, with respect to
the ML detector.

\section{Performance Analysis and Diversity Enhancement}

We investigate the BEP of SS-SIM-OFDM with the ML detection. In particular,
the pairwise error probability (PEP) is analyzed to gain an insight
into the achievable diversity gain and the theoretical bound on the
BEP. More importantly, based on the analysis, we find out two SI mapping
methods which enhance the diversity order of SS-SIM-OFDM.

\subsection{BEP Analysis}

For given $\mathbf{H}$, the well-known conditional PEP of the pairwise
error event (PEE) that the transmitted vector $\mathbf{x}$ is incorrectly
estimated as $\mathbf{\hat{x}}\ne\mathbf{x}$, is given by
\begin{equation}
P\left(\mathbf{x}\rightarrow\mathbf{\hat{x}}|\mathbf{H}\right)=Q\left(\sqrt{\frac{\left\Vert \mathbf{H}\left(\mathbf{v}s-\mathbf{\hat{v}}\hat{s}\right)\right\Vert ^{2}}{2N_{0}}}\right),\label{eq:PEP}
\end{equation}
where we assume that $\mathbf{x}=\mathbf{v}s$ and $\mathbf{\hat{x}}=\mathbf{\hat{v}}\hat{s}$
with $\mathbf{v}=\mathcal{T}\left(\theta,\mathbf{c}\right)$ and $\hat{\mathbf{v}}=\mathcal{T}\left(\hat{\theta},\mathbf{\hat{c}}\right)$,
and $Q\left(.\right)$ denotes the Gaussian tail probability \cite{AlouiniSimon}.
Let us denote $\mathbf{v}=\left[v_{1},...,v_{N}\right]^{T}$, $\mathbf{\hat{v}}=\left[\hat{v}_{1},...,\hat{v}_{N}\right]^{T}$
and $\beta_{n}=\left|v_{n}s-\hat{v}_{n}\hat{s}\right|^{2}$ for $n=1,...,N$.
We rewrite \eqref{eq:PEP} as 
\begin{equation}
P\left(\mathbf{x}\rightarrow\mathbf{\hat{x}}|\mathbf{H}\right)=Q\left(\sqrt{\frac{\sum_{n=1}^{N}\beta_{n}\gamma_{n}}{2}}\right),\label{eq:PEP_rewrite}
\end{equation}
where $\gamma_{n}=\left|h_{n}\right|^{2}/N_{0}=\bar{\gamma}\left|h_{n}\right|^{2}$
is the instantaneous SNR of the $n$-th sub-carrier. Utilizing the
approximation of $Q\left(x\right)\approx e^{-x^{2}/2}/12+e^{-2x^{2}/3}/4$
\cite{AlouiniSimon}, we can approximate \eqref{eq:PEP_rewrite} as
\begin{equation}
P\left(\mathbf{x}\rightarrow\mathbf{\hat{x}}|\mathbf{H}\right)\approx\frac{1}{12}e^{-\frac{\Upsilon}{4}}+\frac{1}{4}e^{-\frac{\Upsilon}{3}},\label{eq:PEP_approx}
\end{equation}
where $\Upsilon=\sum_{n=1}^{N}\beta_{n}\gamma_{n}$. 

Since the Rayleigh channel model is assumed, we attain the moment
generating function (MGF) of $\eta_{n}=\beta_{n}\gamma_{n}$ as $\mathcal{M}_{\eta_{n}}\left(t\right)=\left(1-\beta_{n}\bar{\gamma}t\right)^{-1}$
for $n=1,...,N.$ Consequently, the MGF of $\Upsilon=\sum_{n=1}^{N}\eta_{n}$
is given by $\mathcal{M}_{\Upsilon}\left(t\right)=\prod_{n=1}^{N}\mathcal{M}_{\eta_{n}}\left(t\right)=\prod_{n=1}^{N}\left(1-\beta_{n}\bar{\gamma}t\right)^{-1}.$
Applying the MGF approach to \eqref{eq:PEP_approx}, the unconditional
PEP is obtained as follows\footnote{\textcolor{black}{Note that our analysis here can be applied to any channel model, and hence is not limited to Rayleigh fading. In particular, for a new channel model such as Rician, we just need to apply the MGF of that channel to \eqref{eq:PEP_approx} for achieving the corresponding unconditional PEP. We leave such performance evaluation of our scheme over generalized fading channels as our future work.}}
\begin{equation}
P\left(\mathbf{x}\rightarrow\mathbf{\hat{x}}\right)\approx\frac{1/12}{\prod_{n=1}^{N}\left(1+\frac{\beta_{n}\bar{\gamma}}{4}\right)}+\frac{1/4}{\prod_{n=1}^{N}\left(1+\frac{\beta_{n}\bar{\gamma}}{3}\right)}.\label{eq:uncon_PEP}
\end{equation}

It can be seen from \eqref{eq:uncon_PEP} that the diversity order
of $P\left(\mathbf{x}\rightarrow\mathbf{\hat{x}}\right)$  is the
number of non-zero elements $\beta_{n}$. Thus, letting $d\left(\mathbf{x},\mathbf{\hat{x}}\right)$
be the number of non-zero elements of vector $\mathbf{x}-\mathbf{\hat{x}}$,
the diversity order of SS-SIM-OFDM becomes 
\begin{equation}
G_{d}=\min_{\mathbf{x}\ne\hat{\mathbf{x}}}d\left(\mathbf{x},\mathbf{\hat{x}}\right),\label{eq:G_d}
\end{equation}
which is the minimum value of the diversity orders of all the unconditional
PEPs in \eqref{eq:uncon_PEP}.

To compute $G_{d}$, we investigate two main cases of the PEEs, taking
the proposed rotated ZC codes into account as follows.

\textbf{Case 1} $\theta=\hat{\theta}$: Without loss of generality,
we assume $\theta=\hat{\theta}=\left\{ 1,2,...,K\right\} $, leading
to $\beta_{n}=0$ for $n>K$ and $\beta_{n}=\left|c_{n}s-\hat{c}_{n}\hat{s}\right|^{2}$
for $n=1,...,K$, where it is assumed that $\mathbf{c}=\left[c_{1},...,c_{K}\right]^{T}$
and $\mathbf{\hat{c}}=\left[\hat{c}_{1},...,\hat{c}_{K}\right]^{T}$.
As a result, we deduce $d\left(\mathbf{x},\mathbf{\hat{x}}\right)=d\left(\mathbf{c}s,\mathbf{\hat{c}}\hat{s}\right)\le K$,
which results in three possible cases as follows.

If $s=\hat{s}$ and $\mathbf{c}\ne\hat{\mathbf{c}}$, it is obtained
that $d\left(\mathbf{c}s,\mathbf{\hat{c}}\hat{s}\right)=d\left(\mathbf{c},\mathbf{\hat{c}}\right)$.
In case $K$ is odd, according to \eqref{eq:ZC_sequence}, we assume
that $c_{k}=e^{\varphi_{1}}$ and $c_{k}=e^{\varphi_{2}}$ with $\varphi_{1}=-j\pi d\left(k_{1}^{2}+k_{1}\right)/K+2j\pi l_{1}/B$
and $\varphi_{2}=-j\pi d\left(k_{2}^{2}+k_{2}\right)/K+2j\pi l_{2}/B$,
where $1\le k_{1},k_{2}\le K$ and $0\le l_{1}<l_{2}<K$. Note that
we choose $u=0$ in \eqref{eq:ZC_sequence} to identify such $c_{k}$
and $\hat{c}_{k}$. If $c_{k}=\hat{c}_{k}$, there exists $q\in\mathbb{Z}$
such that $\varphi_{1}=j2\pi q+\varphi_{2}$, which is equivalent
to
\begin{equation}
2K\left(l_{2}-l_{1}\right)=B\left[d\left(k_{2}^{2}+k_{2}-k_{1}^{2}-k_{1}\right)-2qK\right].\label{eq:impossible_1}
\end{equation}
Since $\text{gcd}\left(B,2\right)=\text{gcd}\left(B,K\right)=1$,
we deduce from \eqref{eq:impossible_1} that $l_{2}-l_{1}$ is a multiple
of $B$. However, this clearly conflicts with $0<l_{2}-l_{1}<K<B$.
As a result, $c_{k}\ne\hat{c}_{k}$ for every $k=1,...,K$ and thus
$d\left(\mathbf{x},\mathbf{\hat{x}}\right)=d\left(\mathbf{c},\mathbf{\hat{c}}\right)=K$.
It is worth noting that for even $K$, we arrive at the same conclusion. 

If $s\ne\hat{s}$ and $\mathbf{c}\ne\hat{\mathbf{c}}$, we assume
$s=e^{\frac{j2\pi m_{1}}{M}}$ and $\hat{s}=e^{\frac{j2\pi m_{2}}{M}}$,
$0\le m_{1}<m_{2}<M$, where the PSK modulation is used. If $c_{k}s-\hat{c}_{k}\hat{s}=0,$
similar to the first case, we deduce
\begin{equation}
B\left[d\Phi M-2K\left(qM+m_{2}-m_{1}\right)\right]=2MK\left(l_{2}-l_{1}\right),\label{eq:impossible_2}
\end{equation}
where $\Phi=k_{2}^{2}+k_{2}-k_{1}^{2}-k_{1}$. Due to the fact of
$\text{gcd}\left(B,2\right)=\text{gcd}\left(B,MK\right)=1$, we attain
$l_{2}-l_{1}$ which is a multiple of $B$. This is impossible since
$0<l_{2}-l_{1}<B$. Therefore, $c_{k}s-\hat{c}_{k}\hat{s}\ne0$ for
every $k=1,...,K$, leading to $d\left(\mathbf{x},\mathbf{\hat{x}}\right)=K$.

In case of $s\ne\hat{s}$ and $\mathbf{c}=\hat{\mathbf{c}}$, it is
shown that $\beta_{k}=\left|c_{k}s-\hat{c}_{k}\hat{s}\right|^{2}=\left|c_{k}\right|^{2}\left|s-\hat{s}\right|^{2}\ne0$
for every $k=1,...,K$. Hence, the diversity order achieved in this
case is still $K$.

\textit{Remark 1.} In case of $\theta=\hat{\theta}$, thanks to the
proposed rotated ZC codes, the diversity order of SS-SIM-OFDM is maximized
to be $K$, i.e., the number of active sub-carriers in both the $M$-ary
and SS index domains.

\textbf{Case 2} $\theta\ne\hat{\theta}$: There exists at least two
sub-carrier indices $n_{1}\ne n_{2}\in\left\{ 1,...,N\right\} $ such
that $v_{n_{1}}\ne0$, $\hat{v}_{n_{1}}=0$ and $v_{n_{2}}=0$, $\hat{v}_{n_{2}}\ne0$.
As a consequence, we obtain $\beta_{n_{1}}=\left|v_{n_{1}}s\right|^{2}\ne0$
and $\beta_{n_{2}}=\left|\hat{v}_{n_{2}}\hat{s}\right|^{2}\ne0$,
which leads to $d\left(\mathbf{x},\mathbf{\hat{x}}\right)\ge2.$ However,
employing the conventional SI mapping methods such as the combinatorial
method \cite{basar3013}, there always exist two vectors $\mathbf{x}$
and $\mathbf{\hat{x}}$ satisfying $d\left(\mathbf{x},\mathbf{\hat{x}}\right)=2.$
For example, we consider $\mathbf{x}$ and $\mathbf{\hat{x}}$ that
have $\theta=\left\{ 1,2,...,K\right\} $, $\hat{\theta}=\left\{ 1,2,...,K-1,K+1\right\} $,
$\mathbf{c}=\mathbf{\hat{c}}$, and $s=\hat{s}.$ It is shown from
this example that $\beta_{n}=0$ for $n=1,...,K-1$ and $n=K+2,...,N$,
while $\beta_{K},\beta_{K+1}\ne0$, leading to $d\left(\mathbf{x},\mathbf{\hat{x}}\right)=2.$

\textit{Remark 2:} The diversity order achieved by SS-SIM-OFDM in
case of $\theta\ne\hat{\theta}$ is two, which is far smaller than
that attained in the first case with $\theta=\hat{\theta}$, especially
when $K$ increases. Obviously, the overall diversity order is $G_{d}=2$.
It is also noteworthy that the worst PEE occurs when $\mathbf{x}$
and $\mathbf{\hat{x}}$ have the same spreading code and $M$-ary
symbol, and especially their active sub-carrier indices are identical
at $K-1$ positions. Such insights into the unbalance of the diversity
gains attained in different domains motivate us to propose two methods
to enhance the diversity gain $G_{d}$ in the next section.

Finally, the theoretical upper bound on the BEP of the proposed scheme
is expressed by
\begin{equation}
P_{b}\le\frac{1}{2^{p}p}\sum_{\mathbf{x}}\sum_{\hat{\mathbf{x}}}P\left(\mathbf{x}\rightarrow\mathbf{\hat{x}}\right)w\left(\mathbf{x},\mathbf{\hat{x}}\right),\label{eq:PEP_upper}
\end{equation}
where $P\left(\mathbf{x}\rightarrow\mathbf{\hat{x}}\right)$ is given
in \eqref{eq:uncon_PEP} and $w\left(\mathbf{x},\mathbf{\hat{x}}\right)$
is the number of bit errors of the PEE of $\mathbf{x}\rightarrow\mathbf{\hat{x}}$.

\subsection{Diversity Enhancement}

As analyzed in the previous section, the diversity order of SS-SIM-OFDM
is always limited by two in the SI domain, regardless of increasing
$K$. Moreover, unlike the existing IM schemes, the diversity gain
of the proposed scheme strongly depends on the order of sub-carrier
indices in each SI set $\theta$. For instance, when $(N,K)=(3,2)$,
we consider $\mathcal{I}_{1}=\left\{ \left(1,2\right),\left(2,3\right)\right\} $
and $\mathcal{I}_{2}=\left\{ \left(1,2\right),\left(3,2\right)\right\} $
which have the same elements in every SI set, but in different orders.
The precoding vectors obtained by $\mathcal{I}_{1}$, $\mathcal{I}_{2}$
and $\mathbf{c}=\left[c_{1},c_{2}\right]^{T}$ are $\mathcal{T}\left(\mathcal{I}_{1},\mathbf{c}\right)=\left\{ \left[c_{1},c_{2},0\right]^{T},\left[0,c_{1},c_{2}\right]^{T}\right\} $
and $\mathcal{T}\left(\mathcal{I}_{2},\mathbf{c}\right)=\left\{ \left[c_{1},c_{2},0\right]^{T},\left[0,c_{2},c_{1}\right]^{T}\right\} $.
Because the Hamming distance between two precoding vectors in $\mathcal{T}\left(\mathcal{I}_{1},\mathbf{c}\right)$
is larger than that in $\mathcal{T}\left(\mathcal{I}_{2},\mathbf{c}\right)$,
$\mathcal{I}_{1}$ provides a better diversity gain than $\mathcal{I}_{2}$.
Based on this insight, we propose two diversity enhancement methods
for SS-SIM-OFDM as follows.

\subsubsection{Sub-carrier index set reduction (SISR)}

This simple method is to reduce the number of SI sets $\theta$ used
in the SIM process. Although the SISR method suffers from the loss
of the sub-carrier index bits, we can compensate this loss by increasing
the $M$-ary modulation size. Let us give an example with $(N,K)=(4,3)$.
While the combinatorial method uses all possible 4 SI sets as $\mathcal{I}=\left\{ \left(1,2,3\right),\left(1,2,4\right),\left(1,3,4\right),\left(2,3,4\right)\right\} $,
the proposed SISR can employ either $\mathcal{I}=\left\{ \left(1,2,3\right),\left(2,3,4\right)\right\} $
or $\mathcal{I}=\left\{ \left(1,2,3\right),\left(2,1,4\right)\right\} $
to increase $G_{d}$ from 2 to 3. 

\subsubsection{Ordering Sub-carrier Indices (OSI)}

Since existing IM schemes have the performance independent of the
order of sub-carrier indices in each SI set $\theta$, they just simply
employ the same descending or ascending order for all SI sets. However,
such an ordering is not preferable for SS-SIM-OFDM. Particularly,
the same order not only increases the number of the worst PEEs, but
also limits the achievable diversity order of being two as shown in
Remark 2. Thus, we propose the OSI method to improve the diversity
gain of SS-SIM-OFDM, taking the order of sub-carrier indices into
consideration.

We first derive the criteria for the OSI method in designing $\mathcal{I}=\left\{ \theta_{1},...,\theta_{2^{p_{1}}}\right\} $
to maximize the diversity gain of SS-SIM-OFDM. For given two different
SI sets $\theta_{m}=\left\{ \alpha_{1},...,\alpha_{K}\right\} $ and
$\theta_{n}=\left\{ \mu_{1},...,\mu_{K}\right\} $ in $\mathcal{I}$,
where $m\ne n\in\left\{ 1,...,2^{p_{1}}\right\} $, denote by $\Omega\left(\theta_{m},\theta_{n}\right)$
the number of indices $k\in\left\{ 1,...,K\right\} $ such that $\alpha_{k}\ne\mu_{k}$.
Let $\kappa=\min_{m\ne n}\Omega\left(\theta_{m},\theta_{n}\right)$
and $\Gamma=\sum_{n=1}^{2^{p_{1}}}\sum_{m\ne n=1}^{2^{p_{1}}}\Omega\left(\theta_{m},\theta_{n}\right)$.
The OSI method is to design $\mathcal{I}$ that maximizes $\kappa$
first and then $\Gamma$. It can be seen that $\Omega\left(\theta_{m},\theta_{n}\right)\le K$
for every $m\ne n$, thus $\kappa\le K$ and $\Gamma\le2^{p_{1}}\left(2^{p_{1}}-1\right)K$.
As a result, the best design of $\mathcal{I}$ is the one that satisfies
$\Omega\left(m,n\right)=K$ for every $m\ne n$. 

Based on the criteria above, the OSI method includes two following
steps. The first step is to create $\mathcal{I}_{1}=\left\{ \ddot{\theta}_{1},...,\ddot{\theta}_{2^{p_{1}}}\right\} $
that satisfies the probabilities of indices $i$ appearing in $\mathcal{I}_{1}$
are the same for $i=1,...,N$, without considering the order of sub-carrier
indices. While the equiprobable sub-carrier activation (ESA) algorithm
\cite{ESA2016} can be used to carry out this step, we propose a simpler
approach as follows. Rather than directly designing $\mathcal{I}_{1}$
with $2^{p_{1}}$ SI sets, we just need to design $\mathcal{I}_{2}$
with $T=C\left(N,K\right)-2^{p_{1}}$ SI sets which also have the
same property as $\mathcal{I}_{1}.$ Then, the desired $\mathcal{I}_{1}$
is determined by $\mathcal{I}_{1}=\mathcal{I}_{0}-\mathcal{I}_{2}$,
where $\mathcal{I}_{0}$ contains all $C\left(N,K\right)$ possible
SI sets. Since $T\ll2^{p_{1}}$, designing $\mathcal{I}_{2}$ is easier
than $\mathcal{I}_{1}$. For example, when $\left(N,K\right)=\left(4,2\right)$
with $2^{p_{1}}=4$ and $T=2$, we just simply pick up $\mathcal{I}_{2}=\left\{ \left(1,2\right),\left(3,4\right)\right\} $
to obtain $\mathcal{I}_{1}=\mathcal{I}_{0}-\mathcal{I}_{2}=\left\{ \left(1,3\right),\left(1,4\right),\left(2,3\right),\left(2,4\right)\right\} $.

In the second step, utilizing $\mathcal{I}_{1}$, we design the OSI
algorithm to reorder the indices of every SI set in $\mathcal{I}_{1}$
to attain $\mathcal{I}=\left\{ \theta_{1},...,\theta_{2^{p_{1}}}\right\} $
that fulfils the criteria of maximizing $\kappa$ and $\Gamma$. Notice
that maximizing $\kappa$ is more important than $\Gamma$ since $\kappa$
is the main factor determining the diversity gain in the SI domain.
The proposed OSI algorithm is described in Algorithm 3, where for
each $\ddot{\theta}_{n}\in\mathcal{I}_{1}$, denote by $\mathcal{F}\left(\ddot{\theta}_{n}\right)=\left\{ \ddot{\theta}_{n}^{\left(1\right)},...,\ddot{\theta}_{n}^{\left(L\right)}\right\} $
the set of all $L=K!$ permutations of $\ddot{\theta}_{n}$. To better
understand this algorithm, let us recall the example in the first
step. Particularly, atfter conducting the OSI algorithm with the input
$\mathcal{I}_{1}$, we attain the resulting $\mathcal{I}=\left\{ \left(1,3\right),\left(4,1\right),\left(3,2\right),\left(2,4\right)\right\} $
with $\kappa=K=2$, and $\Gamma=2^{p_{1}}\left(2^{p_{1}}-1\right)K=24$,
while the combinatorial method results in $\kappa=1$ and $\Gamma=16$.
\begin{algorithm}[tbh]
\caption{Ordering Sub-carrier Indices (OSI) Algorithm}
\label{alg:OSI} \textbf{Input:} $\mathcal{I}_{1}=\left\{ \ddot{\theta}_{1},...,\ddot{\theta}_{2^{p_{1}}}\right\} $
obtained from the first step.

\textbf{Output:} $\mathcal{I}=\left\{ \theta_{1},...,\theta_{2^{p_{1}}}\right\} $
\begin{enumerate}
\item Set $\theta_{1}=\ddot{\theta}_{1}$.
\item \textbf{for} $n=2$ \textbf{to} $2^{p_{1}}$ \textbf{do}
\item ~~~Set $\theta_{n}=\ddot{\theta}_{n}^{\left(1\right)}$.
\item \textbf{~~~}$\kappa_{1}=\min_{l=1,...,n-1}\Omega\left(\ddot{\theta}_{n}^{\left(1\right)},\theta_{l}\right)$.
\item ~~~\textbf{for} $m=2$ \textbf{to} $U$ \textbf{do}
\item ~~~~~~$\kappa_{m}=\min_{l=1,...,n-1}\Omega\left(\ddot{\theta}_{n}^{\left(m\right)},\theta_{l}\right)$,
$\ddot{\theta}_{n}^{\left(m\right)}\in\mathcal{F}\left(\ddot{\theta}_{n}\right)$.
\item ~~~~~~$\Gamma_{m}=\sum_{l=1}^{n-1}\Omega\left(\ddot{\theta}_{n}^{\left(m\right)},\theta_{l}\right).$
\item ~~~\textbf{~~~if} $\kappa_{m}>\kappa_{m-1}$ \textbf{or}
\item \textbf{~~~~~~$\kappa_{m}=\kappa_{m-1}$ and $\Gamma_{m}>\Gamma_{m-1}$ }
\item ~~~~~~~~~Set $\theta_{n}=\ddot{\theta}_{n}^{\left(m\right)}.$ 
\item ~~~~~~\textbf{end} \textbf{if}
\item ~~~\textbf{end for}
\item \textbf{end for}
\end{enumerate}
\end{algorithm}

It should be noted that unlike the SISR, the OSI method can significantly
enhance the diversity gain of SS-SIM-OFDM without sacrificing the
SE. This can be shown in Table I, which compares the diversity order
when using the OSI and combinatorial methods for various $\left(N,K\right)$
and $M=2$. We also include the number of the worst PEEs (denoted
by $N_{d}$) whose PEPs have the diversity order of $G_{d}.$ Note
that in case two schemes have the same $G_{d}$, which one has smaller
$N_{d}$ would be better. As seen in Table I, the OSI method provides
a significantly smaller $N_{d}$ and larger $G_{d}$ than the combinatorial
method. Hence, the OSI method can remarkably reduce the BEP of SS-SIM-OFDM
as validated in Section V.

\begin{table}[t]
\caption{$G_{d}$ and $N_{d}$ comparison between the OSI and combinatorial
methods for various values of $(N,K)$ \label{tab:complexity}}

\centering{}%
\begin{tabular}{|c|c|c|c|c|}
\hline 
\multirow{2}{*}{$(N,K)$} & \multicolumn{2}{c|}{OSI} & \multicolumn{2}{c|}{Comb}\tabularnewline
\cline{2-5} \cline{3-5} \cline{4-5} \cline{5-5} 
 & $G_{d}$ & $N_{d}$ & $G_{d}$ & $N_{d}$\tabularnewline
\hline 
$\left(3,2\right)$ & 2 & 6 & 2 & 8\tabularnewline
\hline 
$\left(4,2\right)$ & 2 & 12 & 2 & 20\tabularnewline
\hline 
$\left(5,4\right)$ & 4 & 80 & 2 & 12\tabularnewline
\hline 
$(5,3)$ & 2 & 3 & 2 & 26\tabularnewline
\hline 
\end{tabular}
\end{table}

\section{Simulation Results }

We present simulation results to verify the performance of SS-SIM-OFDM
with various detectors and diversity enhancement methods. We select
SS-OFDM-IM \cite{SSIM2018}, ReMO \cite{ThienTWC2018}, CI-OFDM-IM
\cite{CIbasar2015}, classical OFDM-IM \cite{basar3013} and OFDM
as benchmark schemes. For convenience, the configurations of SS-SIM-OFDM,
ReMO, CI-OFDM-IM and OFDM-IM are represented by $(N,K,M)$, while
that of SS-OFDM-IM is $\left(N,M\right)$, where $M$ is the modulation
size and $N$, $K$ are numbers of sub-carriers and active ones per
cluster, respectively. The PSK modulation is used for all the schemes. \textcolor{black}{We summarize all simulation parameters of the proposed SS-SIM-OFDM scheme in Table~\ref{tab:para}.}

\begin{table}
\caption{\textcolor{black}{A summary of simulation parameters of SS-SIM-OFDM}\label{tab:para}}

\centering{}%
\begin{tabular}{|l|r|}
\hline 
\textcolor{black}{Parameter}  & \textcolor{black}{Value}\tabularnewline
\hline 
\hline 
\textcolor{black}{Number of sub-carriers for each cluster $N$}  & \textcolor{black}{4, 5}\tabularnewline
\hline 
\textcolor{black}{Modulation order $M$}  & \textcolor{black}{2, 4, 8, 16}\tabularnewline
\hline 
\textcolor{black}{Number of active sub-carriers $K$}  & \textcolor{black}{2, 3, 4}\tabularnewline
\hline 
\textcolor{black}{Spectral efficiency (bps/Hz)}  & \textcolor{black}{1, 1.25, 1.5}\tabularnewline
\hline 
\textcolor{black}{Channel model}  & \textcolor{black}{Rayleigh fading}\tabularnewline
\hline 
\textcolor{black}{$M$-ary modulation type}  & \textcolor{black}{PSK}\tabularnewline
\hline 
\textcolor{black}{CSI condition}  & \textcolor{black}{Perfect, imperfect}\tabularnewline
\hline 
\end{tabular}
\end{table}

\begin{figure}[t]
\begin{centering}
\includegraphics[width=1\columnwidth]{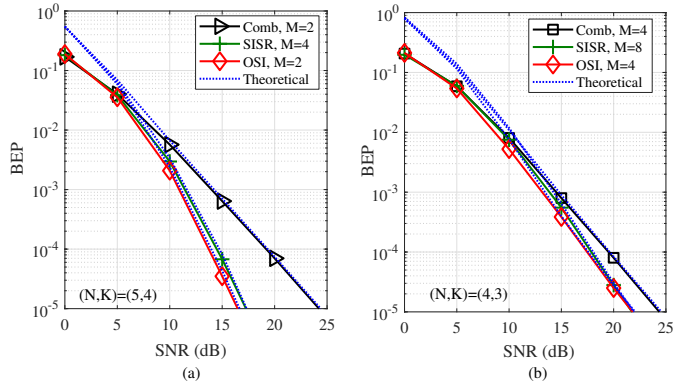} 
\par\end{centering}
\caption{BEP performance of SS-SIM-OFDM when different SI mapping methods are
employed, and (a) $(N,K)=(5,4$), $M=2$ or $M=4$ and (b) $(N,K)=(4,3)$,
$M=4$ or $M=8.$ \label{fig:sim_en}}
\end{figure}
\begin{figure}[t]
\begin{centering}
\includegraphics[width=0.85\columnwidth]{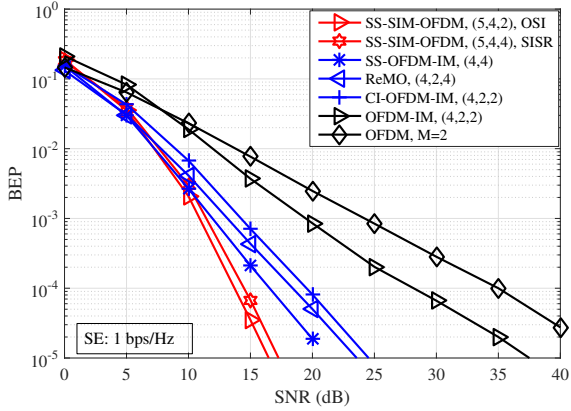}
\par\end{centering}
\caption{BEP comparison among SS-SIM-OFDM and benchmark schemes at the SE of
1 bps/Hz. The ML detection is used. \label{fig:1bit}}
\end{figure}

Fig. \ref{fig:sim_en} depicts the BEP of SS-SIM-OFDM with various
SI mapping methods such as the combinatorial \cite{basar3013}, proposed
SISR and OSI methods. For SISR, we employ a larger $M$ than two remaining
methods to make all of them have the same SE. Particularly, $M=4$
and 8 are used for SISR in Fig. \ref{fig:sim_en}(a) and Fig. \ref{fig:sim_en}(b),
respectively. As seen from Fig. \ref{fig:sim_en}, the proposed SISR
and OSI methods provide better BEP than the combinatorial method,
as both are designed to enhance the diversity gain of SS-SIM-OFDM.
For instance, in Fig. \ref{fig:sim_en}(a), at $\text{BEP}=10^{-4}$,
there are respectively 5 and 5.5 dB SNR gains achieved by SISR and
OSI over the combinatorial method. In addition, the theoretical bound
is very tight, especially at high SNRs. Hence, the derived bound can
be an effective tool to evaluate the BEP of the proposed scheme at
high SNRs. Since OSI performs better than SISR, hereinafter, we will
mainly use OSI in SS-SIM-OFDM for comparisons.

\begin{figure}[t]
\begin{centering}
\includegraphics[width=0.85\columnwidth]{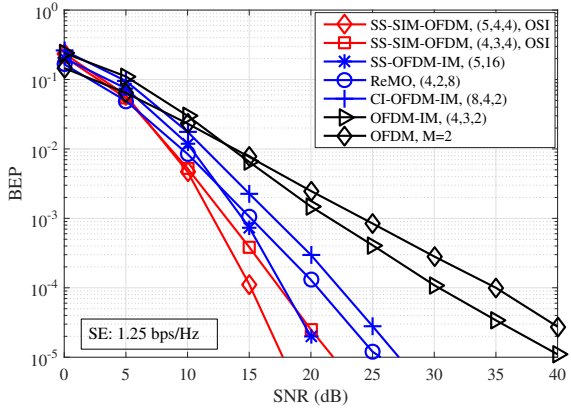}
\par\end{centering}
\caption{BEP comparison among SS-SIM-OFDM and benchmark schemes at the SE of
1.25 bps/Hz. All schemes use the ML detector. \label{fig:1.25bits}}
\end{figure}
\begin{figure}[t]
\begin{centering}
\includegraphics[width=0.85\columnwidth]{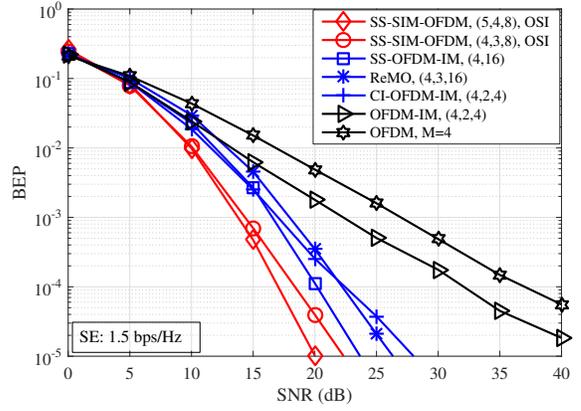}
\par\end{centering}
\caption{BEP comparison among SS-SIM-OFDM and benchmark schemes at the SE of
1.5 bps/Hz and the ML detector is used for all schemes. \label{fig:1.5bits}}
\end{figure}
Fig. \ref{fig:1bit} compares the BEP performance between SS-SIM-OFDM
and benchmark schemes with the ML detector at the SE of 1 bps/Hz.
Obviously, the proposed scheme with either OSI or SISR significantly
outperforms all benchmarks, especially at moderate and high SNRs.
For example, at $\text{BER}=10^{-4}$ in Fig. \ref{fig:1bit}, SS-SIM-OFDM
employing OSI exhibits 3, 5, 6, 15, 21 dB SNR gains over SS-OFDM-IM,
ReMO, CI-OFDM-IM, OFDM-IM and OFDM, respectively. This performance
gain comes from the fact that our scheme is designed specifically
to provide higher diversity gain that the benchmarks, whose gain is
often limited by two. The same observation can be found in Fig. \ref{fig:1.25bits}
with a SE of 1.25 bps/Hz.

In Fig. \ref{fig:1.5bits}, we show the BEP comparison between the
proposed and benchmark schemes at the SE of 1.5 bps/Hz. Once again,
our scheme is superior to all benchmarks in terms of the BEP performance
in a wide range of SNRs. Specifically, at $\text{BEP}=10^{-3}$ in
Fig. \ref{fig:1.5bits}, SS-SIM-OFDM of $(4,3,4)$ outperforms SS-OFDM-IM,
ReMO, CI-OFDM-IM and OFDM-IM with SNR gains of 2, 2.5, 4 and 8 dB,
respectively. Therefore, we can conclude that our SS-SIM-OFDM provides
higher reliability than the benchmark schemes, especially at low SEs. 

\begin{figure}[tb]
\begin{centering}
\includegraphics[width=1\columnwidth]{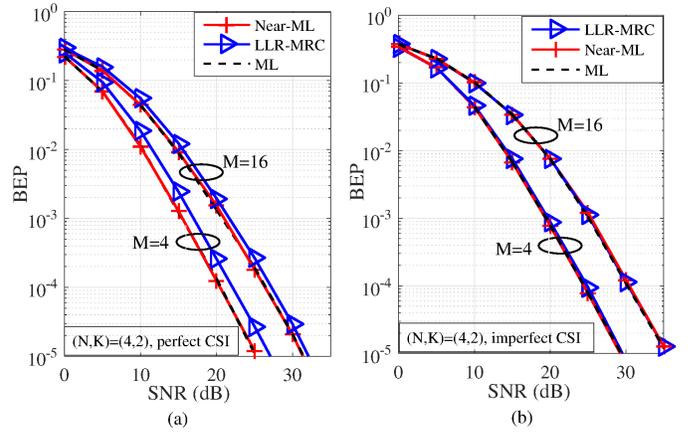} 
\par\end{centering}
\caption{BEP comparison between our near-ML and LLR-MRC detectors and the ML
detector, under (a) perfect and (b) imperfect CSI, when $(N,K)=(4,2)$
and $M=\{4,16\}$. The OSI method is used for our scheme. \label{fig:Detection}}
\end{figure}

Fig. \ref{fig:Detection} presents the BEP comparison between the
proposed near-ML, LLR-MRC detectors and the ML detector when $(N,K)=(4,2)$
and $M=\{4,16\}$, under both perfect and imperfect CSI conditions.
Here, we consider the CSI uncertainty caused by the minimum mean square
error estimator \cite{ThienTVT2017}. \textcolor{black}{It is shown from Fig. \ref{fig:Detection}
that the proposed near-ML achieves a near-optimal BEP as the
ML detector, while the LLR-MRC detector suffers from the performance
loss, particularly when $M$ is small, i.e., $M=4$. This is due to the fact that when the modulation order $M$ is small, the performance of LLR-MRC is more influenced by the index detection error than by the $M$-ary symbol detection error. More particularly, in order to achieve the lowest complexity, LLR-MRC has to detect the indices of active sub-carriers first, which makes it more sensitive to the index detection error than the ML and near-ML detectors that jointly detect both the active indices and $M$-ary symbol. Finally, the gap between LLR-MRC and ML or near-ML becomes
negligible when $M$ increases or the CSI is imperfect, as shown in
Fig. \ref{fig:Detection}(b). As a result, our detectors are
appropriate for practical implementations of SS-SIM-OFDM.}

\section{Conclusion}

We have proposed a novel IM scheme called as SS-SIM-OFDM to improve
the reliability of existing IM-based systems. In SS-SIM-OFM, SS and
SI modulations are jointly employed to form a precoding vector which
is then used to spread an $M$-ary modulated symbol across all active
sub-carriers. As a result, all three possible signal dimensions of
the existing IM schemes, including indices of both spreading codes
and sub-carriers, and the $M$-ary symbol are exploited to convey
data bits. We designed two reduced-complexity detectors, namely the
near-ML and LLR-MRC detectors, whose complexities are roughly independent
of the modulation size. Then, the BEP and its upper bound were derived
to gain an insight into the diversity gain. Based on this, two diversity
enhancement methods, namely SISR and OSI, were proposed, in which
OSI can significantly improve the diversity gain without losing the
SE, while SISR suffers from the decrease of SI bits. Simulation results
showed that our scheme employing SISR and OSI achieves lower BEP than
the benchmarks. In addition, the near-ML detector achieves the optimal
BEP, while the LLR-MRC detector having lowest complexity suffers from
a slight performance loss. However, the performance gap between two
detectors becomes marginal when $M$ increases or the CSI is imperfect.
Our future work will focus on increasing the SE of SS-SIM-OFDM through
transmitting multiple $M$-ary symbols such that it can perform well
at higher SEs.

\bibliographystyle{IEEEtran}
\phantomsection\addcontentsline{toc}{section}{\refname}\bibliography{Ref}

\end{document}